\documentclass{revtex4}

\usepackage{latexsym,amssymb,amsmath}

\usepackage{graphicx}

\def\GeV{{\rm\ GeV}}
\def\ve{\varepsilon}
\def\CG{{\cal G}}

\begin{document}

\title{Phenomenological analysis of two-photon exchange effects\\
 in proton form factor measurements}

\author{Dmitry~Borisyuk}
\email{borisyuk@bitp.kiev.ua} 
\author{Alexander~Kobushkin}%
\email{kobushkin@bitp.kiev.ua}
\affiliation{Bogolyubov Institute for Theoretical Physics\\
Metrologicheskaya street 14-B, 03680, Kiev, Ukraine}

\begin{abstract}We perform a model-independent phenomenological analysis of experimental data on proton form factor ratio. We find that only one of the two-photon exchange amplitudes, which we call $\delta\CG_M$, is responsible for the discrepancy between Rosenbluth and polarization methods. The linearity of Rosenbluth plots implies that $\delta\CG_M$ is approximately linear function of $\ve$. The slope of $\delta\CG_M$ is extracted from experimental data and is shown to be consistent with theoretical calculations.
\end{abstract}

\date{\today}

\maketitle

At present, proton form factors (FFs) are measured with two techniques: Rosenbluth or longitudinal-transverse (LT) separation and polarization transfer (PT). The latter technique gives only FF ratio $G_E/G_M$. When the experiments by the PT method were extended to
$Q^2 \ge 2\GeV^2$, the results have shown clear deviation from those of Rosenbluth method \cite{HowWell}.
Two-photon exchange (TPE) effects are generally believed to be the cause for this discrepancy.
%
They were studied both phenomenologically \cite{GV,Tvaskis,C} and theoretically \cite{Blunden,BMT,BMTdelta,we,GPD}.
The theoretical calculation of TPE effects is nontrivial and technically difficult.
%
On the other hand, assuming that the discrepancy is entirely due to TPE, one can extract some information on TPE directly from experiment.
The values of TPE amplitudes, obtained 
in such a manner
in Ref.\cite{GV}, are in serious disagreement with the theoretical estimates.
In the present paper we extract the TPE amplitudes from experiment
more correctly and demonstrate that they agree with theoretical calculations.

The following notation is used: particle momenta are defined by 
${\rm e}(k) + {\rm p}(p) \to {\rm e}(k') + {\rm p}(p')$; 
$K = (k+k')/2$, $P = (p+p')/2$, $\nu = 4PK$, transferred momentum is
$q = p' - p$, $Q^2 = -q^2$, the proton mass is $M$; the virtual photon polarization parameter is $\ve = \frac{\nu^2 - Q^2(4M^2+Q^2)}{\nu^2 + Q^2(4M^2+Q^2)}$.

In general case the amplitude of the elastic electron-proton scattering  can be written as \cite{GV}
\begin{equation}
 {\cal M} = \frac{4\pi\alpha}{Q^2} \bar u'\gamma_\mu u \cdot
 \bar U' \left(\tilde F_1 \gamma^\mu - \tilde F_2 
 [\gamma^\mu,\gamma^\nu] \frac{q_\nu}{4M} +
 \tilde F_3  K_\nu \gamma^\nu \frac{P^\mu}{M^2}
 \right) U
\end{equation}
(the terms proportional to the electron mass are neglected). Invariant amplitudes (also called generalized FFs) $\tilde F_i$
are scalar functions of two kinematic variables, say, $Q^2$ and $\ve$. 
For our purpose it is convenient to introduce linear combinations
\begin{equation} \label{calG}
 \CG_E = \tilde F_1 - \tau \tilde F_2 + \frac{\nu}{4M^2} \tilde F_3, \ \ \ 
 \CG_M = \tilde F_1 +\tilde F_2 + \ve \frac{\nu}{4M^2} \tilde F_3, \ \ \ 
\end{equation}
where $\tau = \frac{Q^2}{4M^2}$. In general, the amplitudes are complex, but 
their imaginary parts give negligible contribution to the observables considered here.
Thus later on, speaking of the amplitudes, we understand their real parts.

With amplitudes (\ref{calG}), the reduced cross-section for unpolarized particles can be written like the Rosenbluth formula
\begin{equation}
 \frac{(1-\ve)}{32\alpha^2} Q^2 \left(\frac{\nu+Q^2}{\nu-Q^2} \right)^2 \frac{d\sigma}{d\Omega_{lab}} \equiv \sigma_R = \ve \CG_E^2 + \tau \CG_M^2 + O(\alpha^2). 
\end{equation}
%
The amplitudes can be decomposed as
\begin{equation}
 \CG_E(Q^2,\ve) = G_E(Q^2) + \delta G_E^{(T)} (Q^2,\ve) + \delta\CG_E(Q^2,\ve) + O(\alpha^2)
\end{equation}
and similarly for $\CG_M$; $G_E$ and $G_M$ are proton electric and magnetic FFs
and $\delta G_{E,M}^{(T)} + \delta\CG_{E,M}$ are TPE corrections of order $\alpha$.
Here $\delta G_{E,M}^{(T)}$ denotes the part of the correction, calculated by Tsai \cite{Tsai}.
%
%
All infrared divergence is contained in it. 
The reduced cross-section then takes the form
\begin{equation}
 \sigma_R = \ve G_E^2 + \tau G_M^2 + 2 \ve G_E \delta\CG_E + 2 \tau G_M \delta\CG_M + 2 \ve G_E \delta G_E^{(T)} + 2 \tau G_M \delta G_M^{(T)} + O(\alpha^2).
\end{equation}
The terms containing $\delta G_{E,M}^{(T)}$  are always subtracted from the cross-section by experimenters as a part of radiative corrections, so published cross-sections are, dropping terms of order $\alpha^2$,
\begin{equation} \label{Rosen}
 \sigma_R = \ve G_E^2 + \tau G_M^2 + 2 \ve G_E \delta\CG_E + 2 \tau G_M \delta\CG_M.
\end{equation}
These cross-sections are used as an input for the Rosenbluth separation.

Now we will analyze Eq.(\ref{Rosen}) in the case of $\tau \gtrsim 1$.

Since the TPE amplitudes are proportional to $\alpha \approx \frac{1}{137}$, it is natural to believe that their magnitude is about 1\%.
The key point is that $G_M$ is enhanced with respect to $G_E$ by about a factor of $\mu \approx 3$ (proton magnetic moment), and even more according to PT data, so we have
\begin{equation}
 \tau G_M^2 \gg \ve G_E^2 \gg 2 \ve G_E \delta\CG_E, \ \ \tau G_M^2 \gg 2 \tau G_M \delta\CG_M \gg 2 \ve G_E \delta\CG_E.
\end{equation}
Therefore the term $2 \ve G_E \delta\CG_E$ is much smaller than three other terms and can be safely neglected. Instead, the term $2 \tau G_M \delta\CG_M$ can be comparable with $\ve G_E^2$ and thus strongly affect the results of Rosenbluth separation. 
We take for granted that Rosenbluth plots are linear (recent analyses \cite{Tvaskis,C} confirm this),
i.e. $\sigma_R$ is linear function of $\ve$. This necessarily leads to the conclusion, that $\delta \CG_M$ is also approximately linear; we write it as
\begin{equation} \label{linear}
 \delta \CG_M(Q^2,\ve) = [a(Q^2) + \ve b(Q^2)] G_M(Q^2),
\end{equation}
where $a$ and $b$ are of order $\alpha$. 
The coefficient $a$ yields only a small contribution to the large $\ve$-independent term in $\sigma_R$ and does not change $\ve$-dependent term, thus we also may neglect it.
After that
\begin{equation}
 \sigma_R = \tau G_M^2 + \ve (G_E^2 + 2 \tau b \, G_M^2 )
\end{equation}
and the FF ratio squared obtained by the Rosenbluth method is actually
\begin{equation} \label{R_LT}
\biggl. \left(\frac{G_E}{G_M} \right)^2 \biggr|_{LT} \equiv R_{LT}^2 = \frac{G_E^2}{G_M^2} + 2 \tau b.
\end{equation}
The FF ratio obtained in the PT experiments (with the inclusion of TPE) can be written as
\begin{equation}
 \left. \frac{G_E}{G_M} \right|_{PT} \equiv R_{PT} =
 \frac{\CG_E}{\CG_M} \left( 1 - \frac{\ve(1-\ve)}{1+\ve} Y_{2\gamma} \right)
 + O(\alpha^2),
\end{equation}
where $Y_{2\gamma} = \frac{\nu}{4M^2} \frac{\tilde F_3}{G_M}$. The quantity $\frac{\ve(1-\ve)}{1+\ve}$ never exceeds 0.17 and $Y_{2\gamma}$ should be about 1\%, thus the bracket is very close to 1. Also, since the corrections $\delta \CG_E$ and $\delta \CG_M$ amount to about 1\%, the ratio $\CG_E/\CG_M$ is close to $G_E/G_M$ and anyway matches it within the experimental errors, which are typically about 5\%.
Therefore we may say that PT experiments yield {\it true} FF ratio
\begin{equation} \label{R_P}
 R_{PT} = \frac{G_E}{G_M}.
\end{equation}
Substituting (\ref{R_P}) into (\ref{R_LT}), we obtain the TPE correction slope
\begin{equation}
 b = \frac{1}{2\tau}(R_{LT}^2 - R_{PT}^2).
\end{equation}

\begin{figure}[h]
\centering
\includegraphics[width=0.5\textwidth]{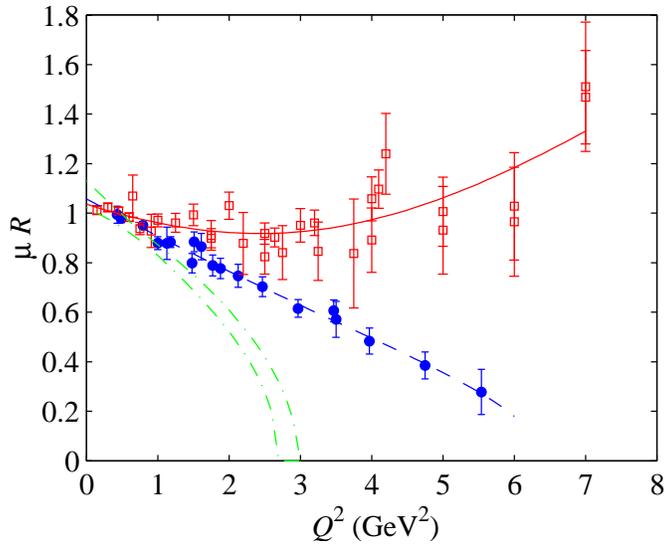}
\caption{(Color online) Experimental data and fits to FF ratio obtained by Rosenbluth method ($\mu R_{LT}$, solid curve and squares) and by PT method ($\mu R_{PT}$, dashed curve and circles) and prediction for Rosenbluth method with positrons (dash-dotted curves, $1\sigma$ range).}
\label{expdata}
\end{figure}

\begin{figure}[h]
\centering
\includegraphics[width=0.45\textwidth]{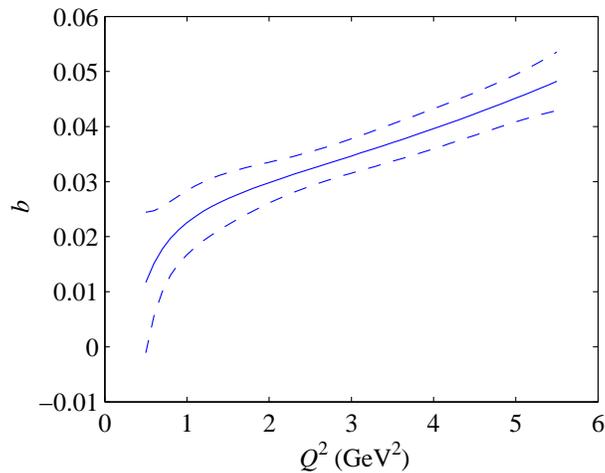}
\caption{(Color online) Extracted TPE correction slope $b(Q^2)$. The dashed curves indicate estimated errors.}\label{coefb}
\end{figure}

We fitted $R_{PT}^2$ from Ref.\cite{PT} and $R_{LT}^2$ from Ref.\cite{LT} with polynomials in $Q^2$ of the third order to minimize the $\chi^2$ function
\begin{equation}
 \chi^2 = \sum_i \left(\frac{R_i^2 - P(Q^2_i)}{2 R_i \Delta R_i} \right)^2,
\end{equation}
where $\Delta R_i$ are experimental errors. We have obtained, for Rosenbluth data
\begin{equation}
 P_{LT}(Q^2) = \mu^{-2}(1.0736 - 0.1864 Q^2 + 0.0358 Q^4 + 0.0007 Q^6)
\end{equation}
with $\chi^2 = 24.2$ for 28 d.o.f. and for PT data
\begin{equation}
 P_{PT}(Q^2) =\mu^{-2}(1.1184 - 0.3256 Q^2 + 0.0323 Q^4 - 0.0014 Q^6)
\end{equation}
with $\chi^2 = 6.6$ for 15 d.o.f. (everywhere $Q$ is in units of GeV and $\mu=2.793$). The results are shown in Fig.\ref{expdata}.
From this we derive the coefficient $b$:
\begin{equation}
 b = - 0.0101/Q^2 + 0.0314 + 0.0008 Q^2 + 0.0005 Q^4.
\end{equation}
%
The dependence of $b$ vs. $Q^2$ is displayed in Fig.\ref{coefb}. The coefficient $a$ cannot be derived from the experimental data.
We see that $b$ varies from 0.01 to 0.05; if the unknown coefficient $a$ is negative and equals about $-b/2$ then the relative size of TPE correction $\delta \CG_M/G_M = a + \ve b$ will never exceed 0.025, in consistency with the assumptions made at the very beginning.

\begin{figure}[h]
\centering
\includegraphics[width=0.45\textwidth]{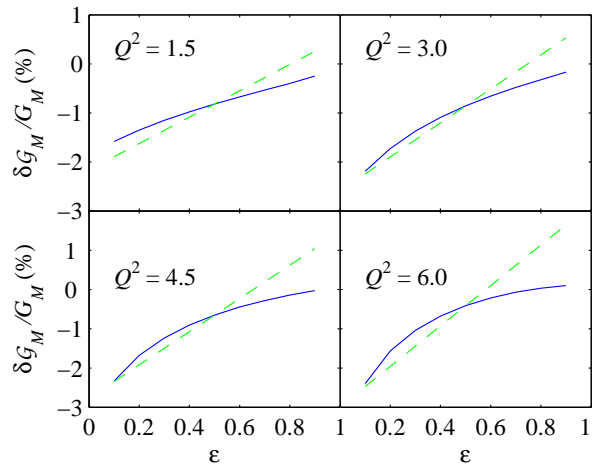}
\caption{(Color online) Comparison of extracted (dashed lines) and calculated (solid curves) values of TPE amplitude $\delta\CG_M/G_M$.}\label{ampls}
\end{figure}

From the theoretical point of view, the TPE corrections can be split into elastic and inelastic parts. 
The elastic part was first calculated in Ref.~\cite{Blunden} using simple
monopole parameterization of FFs and then in a model-independent way in Refs.~\cite{BMT,we}.
The inelastic part is at present modeled by resonances \cite{BMTdelta}
or by parton models \cite{GPD}.
In Fig.\ref{ampls} we plot the elastic part of $\delta \CG_M$, calculated as described in Ref.\cite{we} and straight line according to Eq.(\ref{linear}) with $b$ given above and $a$ chosen arbitrarily. A qualitative agreement is seen at all values of $Q^2$. The gap between the curve and the line is small and should be possibly attributed to inelastic contribution. 
This results differ from those of Ref.\cite{GV}, where authors, in fact, made three arbitrary assumptions about the TPE amplitudes (namely, in the notation of Ref.\cite{GV}, that $Y_{2\gamma}$, $|\tilde G_E|$, and $|\tilde G_M|$ are all independent of $\ve$), and obtained the quantity $Y_{2\gamma}$ approximately 5 times larger than theoretical estimates \cite{BMT}.

If the positrons are used instead of electrons, the TPE corrections change their sign. Thus we have for the Rosenbluth FF ratio squared, measured in positron-proton scattering
\begin{equation}
 \tilde R_{LT}^2 = \frac{G_E^2}{G_M^2} - 2 \tau b 
\end{equation}
(Fig.\ref{expdata}, where dash-dotted lines indicate estimated $1\sigma$ bounds for $\mu\tilde R_{LT}$). The values of $\tilde R_{LT}^2$ strikingly differ from $R_{LT}^2$ and become even negative for $Q^2 > 3 \GeV^2$. Therefore the Rosenbluth measurements with positron beam at $Q^2 \gtrsim 2\GeV^2$ would be an interesting test for TPE effects, and, on the other hand, a new way to measure true FF ratio, since the TPE corrections fully cancel in the average of $R_{LT}^2$ and $\tilde R_{LT}^2$.
Precision study of positron-proton cross-section is already proposed, for $Q^2 \lesssim 1.6\GeV^2$ \cite{VEPP} and for $0.5 < Q^2 < 3 \GeV^2$ \cite{E04}.  
\begin{acknowledgements}
This work was supported by Program of Fundamental Research of the Department of Physics and Astronomy of National Academy of Sciences of Ukraine.
\end{acknowledgements}

\end{document}